\documentclass[aps,pra,showpacs,groupedaddress,twocolumn]{revtex4-1}

\usepackage{graphicx}
\usepackage[ansinew]{inputenc}
\usepackage{array}
\usepackage{color}
\usepackage{amsmath}
\usepackage{amsxtra}
\usepackage{amstext}
\usepackage{amssymb}
\usepackage{latexsym}
\usepackage{dsfont}
\usepackage{verbatim}
\begin{document}

\title{Generation of mechanical squeezing via magnetic dipoles on cantilevers}

\author{H. Seok, L. F. Buchmann, S. Singh, S. K. Steinke, and P. Meystre}
\affiliation{B2 Institute, Department of Physics and College of Optical
Sciences\\The University of Arizona, Tucson, Arizona 85721}
\date{\today}

\begin{abstract}
A scheme to squeeze the center-of-mass motional quadratures of a quantum mechanical oscillator below its standard quantum limit is proposed and analyzed theoretically. It relies on the dipole-dipole coupling between a magnetic dipole mounted on the tip of a cantilever to equally oriented dipoles located on a mesoscopic tuning fork.  We also investigate the influence of several sources of noise on the achievable squeezing, including classical noise in the driving fork and the clamping noise in the oscillator. A detection of the state of the cantilever based on state transfer to a light field is considered. We investigate possible limitations of that scheme.

\end{abstract}

\maketitle

\section{Introduction}

Mechanical cantilevers have a long and rich history as force and field meters. In recent decades, microfabrication and nanotechnology have resulted in spectacular advances with the development of tip microscopy \cite{BinnigRohrer}, resulting in the measurement of feeble forces and the imaging of single atoms, nanoscale magnetic resonance imaging \cite{RugarPNAS}, and single spin detection \cite{SingleSpinDet}, to mention just a few applications. With recent progress in cavity optomechanics and the successful cooling of at least three micromechanical systems \cite{cool1,cool2,cool3} to the deep quantum regime with just a fraction of a phonon of center-of-mass excitation left, mechanical sensing is at the threshold of an important new breakthrough.

One potential advantage of operating micromechanical sensors in the quantum regime is that this opens the way to measurement techniques that can circumvent the standard quantum limit. These techniques rely on the capability to locate the unavoidable quantum noise in a quadrature of the field to be measured that does not interact with the measuring apparatus, and to implement back-action evading techniques that prevent that noise to feed back into the outcome of successive measurements. In quantum optics, the most famous states that permit to achieve this goal are single-mode squeezed states, where the variance of one of the quadratures $\hat X=(\hat a+\hat a^\dagger)/2$ or $\hat X_2=(\hat a-\hat a^\dagger)/2i$ of the light field is below 1/4,  with $\langle \Delta \hat X_1^2 \Delta \hat X_2^2\rangle \ge 1/16$. Here $\hat a$ and $\hat a^\dagger$ are normalized bosonic annihilation and creation operators of the field mode, and $\langle \Delta \hat X_i^2\rangle = \langle \hat X_i^2\rangle - \langle  \hat X_i \rangle ^2$.

While the classical noise squeezing of micromechanical oscillators has already been achieved \cite{rugar,schwab2}, squeezing below the standard quantum limit still remains to be demonstrated. A number of techniques have been proposed, including conditional squeezing using a parametrically coupled electromagnetic cavity driven by one \cite{schwab} or two sidebands \cite{braginsky,clerk} detuned from the cavity resonance by the mechanical oscillation frequency, and are expected to be demonstrated experimentally in the near future. This paper proposes an alternative scheme where a nanoscale cantilever can be prepared in a squeezed state by purely mechanical means via the nonlinearity provided by the magnetic dipole-dipole interaction between a vibrating classical fork and the cantilever. We also show that the squeezing can be detected via state transfer to an optical field coupled to the cantilever in an optical resonator configuration~\cite{Braunstein,TianWang,Khalili}. We discuss the impact of the various sources of noise, including both the noise of the classical driving force and the clamping noise of the cantilever and also comment on limitations to the state transfer scheme resulting from the dynamics of the light field and from the opening of an associated dissipation channel.

The remainder of this paper is organized as follows: Section II introduces our model system and demonstrates how the anharmonic potential that describes the magnetic coupling between the classical fork and the quantum mechanical cantilever results in quadrature squeezing of the cantilever motion under appropriate conditions. Section III analyzes the robustness of the system against various sources of noise. Section IV discusses the beam-splitter state transfer mechanism, and section V is a conclusion and outlook. Calculational details are relegated to two appendices.

\begin{figure}[t]
\includegraphics[width=8 cm]{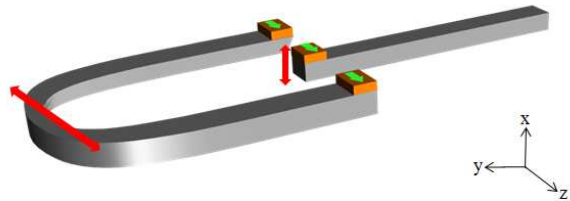}
\caption{\label{fig:setup} Classical tuning fork magnetically coupled to a quantum-mechanical cantilever. The equilibrium position of the cantilever is equidistant from the two extremities of the nanoscale tuning fork.}
\end{figure}

\section{Model system}
We consider a nanomechanical system consisting of a cantilever magnetically coupled to a classical nanoscale tuning fork, as shown in Fig.~\ref{fig:setup}, the coupling being realized via point-like magnetic dipoles located at both extremities of the fork as well as on the cantilever. The oscillation direction of the cantilever and the fork are taken to be the $x$-axis and $z$-axis, respectively, and the magnetic dipoles are assumed to point in the positive $z$-direction. We consider specifically the center-of-mass mode of vibration of the fork, in which the distance between its two extremities remains constant. This arrangement provides a stable mode of operation for experimentally reasonable parameters. 

Denoting the displacement of the cantilever from its equilibrium position by $x_c$, the interaction energy between the three magnetic dipoles can be written as
\begin{equation}
V= \frac{\mu_0}{4 \pi} d_{f}d_{c} \left[\frac{x_c^{2}-2 \ell_+^{2}}{\left(x_c^{2}+\ell_+^{2}\right)^{5/2}}+ \frac{x_c^{2}-2 \ell_-^{2}}{\left(x_c^{2}+\ell_-^{2}\right)^{5/2}}\right],\label{fullinteraction}
\end{equation}
where $d_f$ and $d_c$ are the dipoles moments of the magnets attached to the tips of the fork and to the cantilever, respectively, and $\ell_\pm$ are the distances  between the two tips of the fork and the cantilever along the $z$-direction.

\subsection{Symmetric case}

We assume first that the equilibrium position of the nanomagnet on the cantilever is equidistant from the two tips of the fork. For that setup the attractive forces from the two dipoles acting on the cantilever cancel each other, but for any departure from that situation this is no longer the case, so that the stability of the system must be enforced by the stiffness of the cantilever.

For the mode of vibration under consideration we have in the symmetric case
\begin{subequations}\label{Lplusminusdef}
\begin{eqnarray}
\ell_+ &\approx& \ell + z_f, \\
\ell_- &\approx& \ell - z_f,
\end{eqnarray}
\end{subequations}
where $z_f$ denotes the displacement of the tuning fork tips from their equilibrium position and $\ell = (\ell_+ + \ell_-)/2$. For small displacements of the mechanical elements, $z_f, x_c \ll \ell$, the interaction Hamiltonian can be expanded to second order in $x_c$, yielding
\begin{equation}
V \approx \frac{\mu_0}{4 \pi} \frac{d_{f}d_{c}}{\ell^5} \left[-24z_f^2 +12x_c^2+180\frac{z_f^2x_c^2}{\ell^2}\right],\label{interactionenergy}
\end{equation}
where we ignored a constant term. The first two terms in Eq.~(\ref{interactionenergy}) describe frequency shifts due to the magnetic interaction.

For high amplitude driving, the motion of the fork of effective mass $m_f$ and frequency $\omega_f$ can be approximated by its classical amplitude. The cantilever motion, on the other hand, is treated quantum mechanically, with its displacement $x_c$ given in terms of the bosonic annihilation and creation operators $\hat b$ and $\hat b^\dagger$ by
\begin{equation}
{\hat x}_c=x_0(\hat{b}+\hat{b}^{\dagger}),
\end{equation}
with
\begin{equation}
x_0 = \sqrt{\frac{\hbar}{2m_c\omega_c}},
\end{equation}
where $\omega_c$ and $m_c$ denote the cantilever's frequency and effective mass. The Hamiltonian governing the dynamics of the cantilever is then
\begin{equation}
H=\hbar \omega_c' \hat{b}^{\dagger}\hat{b}+\hat{V},
\end{equation}
where
\begin{equation}
\omega_c^{\prime}= \omega_c + \Delta\omega_c,
\end{equation}
with $\Delta\omega_c$ the frequency shift from the magnetic interaction,
\begin{equation}
\Delta\omega_c \approx \frac{6\mu_0 d_f d_c}{\hbar \pi}\frac{x_0^2}{\ell^5},
\end{equation}
see Eq.~(\ref{interactionenergy}). In terms of $\hat{b}$ and $\hat{b}^{\dagger}$, the dipole interaction $\hat V$ becomes
\begin{equation}
V= \frac{45 \mu_0 d_f d_c}{\pi}\frac{z_0^2x_0^2}{\ell^7}(\beta+\beta^{*})^2(\hat{b}+\hat{b}^{\dagger})^{2},
\label{V1}
\end{equation}
where 
\begin{equation}
\beta = \frac{z_f}{2z_0} + i\frac{z_0p_f}{\hbar},
\end{equation}
and 
\begin{eqnarray}
z_0 &=& \sqrt{\frac{\hbar}{2m_f\omega_f}}.
\end{eqnarray}
A final simplification is obtained by taking the driving frequency of the fork to be
\begin{equation}
\omega_f = \omega_c^{\prime},
\end{equation}
then invoking the rotating wave approximation and switching to a frame rotating with $\omega_f$, where $V$ reduces to the familiar single-mode squeezing Hamiltonian (see e.g.~\cite{Scully2008})
\begin{equation}\label{squeezinghamiltonian}
V=\frac{1}{2}\hbar\chi_s\left(\hat{b}^{\dagger 2}e^{-i2\phi} + \hat{b}^2e^{i2\phi}\right)
\end{equation}
with
\begin{equation}
\chi_s=\frac{90 \mu_0 d_f d_c}{\hbar \pi} \frac{ z_0^2x_0^2}{\ell^7}|\beta|^2
\label{chis}
\end{equation}
and  $\phi$ the relative phase between the classical fork and the cantilever. It defines which quadrature gets squeezed, and in the following we always consider the quadrature which experiences maximum squeezing, omitting the explicit value of $\phi$. 

\subsection{Asymmetric case}

We now turn to the case where the equilibrium position of the cantilever nanomagnet is displaced by a distance $\zeta$ from the center of the fork, the stability of that configuration being guaranteed as before by the mechanical stiffness of the cantilever. We now have
\begin{subequations}\label{Lplusminusdef2}
\begin{eqnarray}
\ell_+ &\approx& \ell + \zeta +z_f, \\
\ell_- &\approx& \ell - \zeta - z_f,
\end{eqnarray}
\end{subequations}
and within the same limit as before
\begin{equation}\label{deltainteractionenergy}
V \approx \frac{\mu_0}{4 \pi} \frac{d_{f}d_{c}}{\ell^5} \left[-24(\zeta+z_f)^2+12x_c^{2} +180\frac{(\zeta+z_f)^2x_c^{2}}{\ell^2}\right].
\end{equation}
When compared to the symmetric case, Eq. (\ref{interactionenergy}), the interaction~(\ref{deltainteractionenergy}) comprises an additional squeezing contribution given by the term proportional to  $\zeta z_f x_c^2$. By driving at twice the cantilever frequency, and invoking the rotating wave approximation, one can access either of the two squeezing interactions separately, in a fashion reminiscent of the situation in parametrically coupled optomechanical resonators \cite{schwab}.

The resulting Hamiltonian is then the sum of two terms of the same form as in Eq.~(\ref{squeezinghamiltonian}) with the coupling coefficient of the term oscillating at  $2\omega_c'$ given by
\begin{equation}
\chi_2(\Delta)=\frac{15\mu_0 d_f d_c}{\hbar \pi}\left[\frac{1}{(\ell-\zeta)^6}-\frac{1}{(\ell+\zeta)^6}\right]x_0^2z_0|\beta|,
\label{chi2}
\end{equation}
and that for the term oscillating at $ \omega_c^{\prime}$ by
\begin{equation}
\chi_1(\zeta)=\frac{45 \mu_0 d_f d_c}{\hbar \pi} \left[\frac{1}{(\ell-\zeta)^7}+\frac{1}{(\ell+\zeta)^7}\right] x_0^2z_0^2|\beta|^2.
\label{chid}
\end{equation}
The coupling constant $\chi_1(\zeta)$ scales as the square of the amplitude of oscillations $|\beta|^2$ of the classical fork, rather than $|\beta|$ as is the case for $\chi_2(\zeta)$, and hence can be dominant for strong fork driving. However, its dependence on $\ell^{-7}$ rather than $\ell^{-6}$ indicates that for appropriate values of $\ell$ the term proportional to $\chi_2$ can be dominant instead, see Fig.~2.

\subsection{Experimental considerations}

Consider for concreteness a nanomechanical cantilever with natural frequency $\omega_c = 2\pi \times 168$ $\mathrm{kHz}$, effective mass $m_{c} = 5 \times 10^{-15}$ $\mathrm{kg}$, and magnetic dipole moment $d_{c}= 2\times 10^{-14}$ $\mathrm{A\cdot m^2}$. The fork is assumed to have effective mass $m_{f} = 1 \times 10^{-11}$ $\mathrm{kg}$, tips separation $2\ell = 2.0$ $\mathrm{\mu m}$, and magnetic dipole moments $d_{f}= 1 \times 10^{-10}$ $\mathrm{A\cdot m^2}$.  These values result in a frequency shift of the cantilever of $\Delta\omega_c \approx 2\pi \times 72$ $\mathrm{MHz}\gg\omega_c$, indicating that the driving frequency of the fork is dominated by the frequency shift  $\omega_f \approx \Delta\omega_c$.  We further assume a fork oscillation amplitude of $A_f = 2z_0|\beta| = 10$ $\mathrm{nm}$, i.e. $A_f=0.01\ell$, with associated mean phonon occupation $|\beta|^{2} = 2.1 \times 10^{15}$, which justifies a classical treatment.

\begin{figure}
\includegraphics[width=0.48\textwidth]{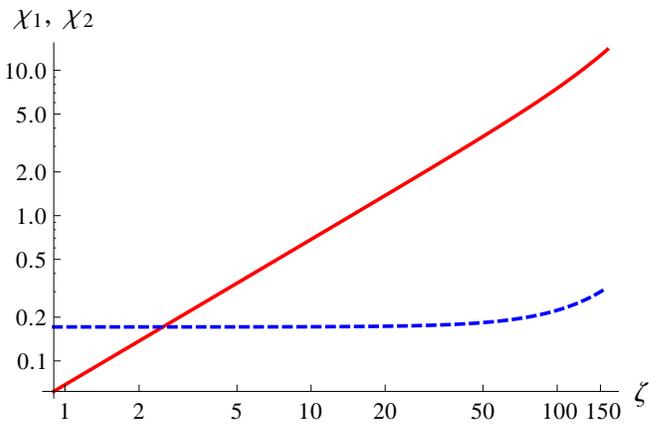}
\caption{(Color Online) Coupling frequencies $\chi_2(\zeta)$~(red, solid), and $\chi_1(\zeta)$(blue, dashed) in MHz as a function of displacement $\zeta$ in nm for parameters of section II.C.}~
\label{chiomega}
\end{figure}

For the asymmetric setup, we need to estimate the frequency of the cantilever in the $z$-direction, which we obtain from $\omega_c^{(z)}/\omega_c\approx w/h$, where $h$ and $w$ are the cantilever's thickness and width. For $h = 12$ nm and $w=600$nm this gives $\omega_c^{(z)}\approx 2\pi \times 8.4 \mathrm{MHz}$. With this value and Eqs.~(\ref{Lplusminusdef2}), we can find the critical points of the full interaction potential, Eq.~(\ref{fullinteraction}). For these  parameters the maximum value of $\zeta$ that yields a stable configuration is $162\mathrm{nm}$. The resulting strengths of the squeezing interaction are shown in Fig.~\ref{chiomega} as a function of  $\zeta$. 

\section{Fluctuations}

It is known that in the absence of fluctuations and starting from the oscillator ground state the squeezing Hamiltonian~(\ref{squeezinghamiltonian}) produces a perfect ``squeezed vacuum'' with average phonon number
\begin{equation}
\langle\hat{b}^\dag\hat{b}\rangle=\sinh^2(\chi t).
\end{equation}
(In the following we use generically the symbol $\chi$ to describe all setups of the previous section.) Clearly, this is unrealistic. What is missing from the discussion so far is a proper accounting of fluctuations. 
This section discusses the effects of three sources of technical noise: the amplitude and phase fluctuations of the fork motion and the clamping noise resulting from the attachment of the cantilever to a thermal reservoir.

\subsection{Amplitude fluctuations}
We assume that the cantilever is slightly displaced from the center of the fork, driven at frequency $\omega_f=2\omega_c'$, so that the squeezing strength is given by Eq.~(\ref{chi2}). The other two cases can be handled similarly and we give the results pertaining to them at the end of this subsection. Assuming as usual that the amplitude of oscillations of the tuning fork can be decomposed into the sum of a constant amplitude and random amplitude fluctuations 
\begin{equation}
|\beta| (t) = |\beta_0|+ \delta\beta(t),
\end{equation}
we find the equation of motion
\begin{equation}
\dot{\hat{b}} = -\left[\chi_0+\delta\chi(t)\right]\hat{b}^{\dagger},
\end{equation}
where $\chi_0$ is given by Eq.~(\ref{chi2}) and
\begin{equation}\label{deltachi}
\delta\chi(t)=\frac{\chi_0}{|\beta_0|}\delta\beta(t).
\end{equation}
Following Ref.~\cite{Zubairy1983}, we assume that the mean and two-time correlation functions of the amplitude fluctuations are described by an Ornstein-Uehlenbeck process, so that
\begin{eqnarray}
\langle\delta\chi(t)\rangle &=& 0, \\
\langle\delta\chi(t)\delta\chi(t^\prime)\rangle &=& \frac{\chi_0^2}{|\beta_0|^2}\sigma\Gamma e^{-\Gamma |t-t^\prime|}\label{chicorrel},
\end{eqnarray}
where $\sigma$ is proportional to the variance of amplitude fluctuations and $\Gamma$ is the inverse of the correlation time of the fluctuations. From Ref.~\cite{Zubairy1983}, we find readily that the variances of the quadratures $\hat X_1$ and $\hat X_2$ of the phonon mode are then given by
\begin{eqnarray}
(\Delta\hat{X}_1(t))^2 &=& e^{-2\chi_0 t}e^{-2{\int_0^t dt' \delta\chi(t^{\prime})}}(\Delta\hat{X}_1(0))^2, \\
(\Delta\hat{X}_2(t))^2 &=& e^{2\chi_0 t}e^{2{\int_0^t dt' \delta\chi(t^{\prime})}}(\Delta\hat{X}_2(0))^2,
\end{eqnarray}
which, upon carrying out the appropriate ensemble averages and taking into account that the initial state of the cantilever is uncorrelated with the fluctuations, yields
\begin{eqnarray}
(\Delta \hat{X}_1(t))^2 &=& \frac{1}{4}e^{-2\chi_0t +4f(t)}, \\
(\Delta \hat{X}_2(t))^2 &=& \frac{1}{4}e^{2\chi_0t +4f(t)},
\end{eqnarray}
where
\begin{equation}
f(t) = \frac{\chi_0^2}{|\beta_0|^2}\frac{\sigma}{\Gamma}\left[e^{-\Gamma t}+\Gamma t-1\right].\label{foft}
\end{equation}

Amplitude fluctuations reduce the rate at which the variances get squeezed. They do not however limit  the maximum squeezing that can be generated. Moreover, Eq.~(\ref{foft}) shows that the lengthening in timescale resulting from amplitude fluctuations scales as $\chi_0^2/|\beta_0|^2$, which is typically a small factor, indicating that in contrast to phase fluctuations to which we turn next, amplitude fluctuations in the classical drive do not significantly affect the dynamics of the system. Finally, we note that if the squeezing factor is given by Eq. (\ref{chid}), we have to modify Eq.~(\ref{deltachi}) with an additional factor of 2, and the function $f(t)$ of Eq.~(\ref{foft}) is thus multiplied by a factor of 4.
\vspace{12pt}

\subsection{Phase fluctuations}
\begin{figure}[t]
\includegraphics[width=0.48 \textwidth]{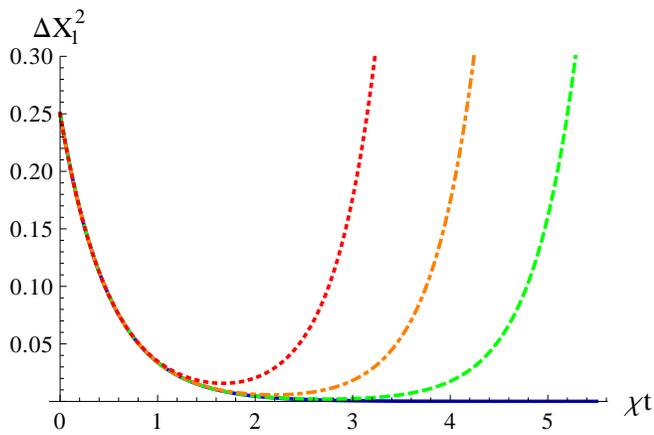}
\caption{\label{fig:phase} $(\Delta X_1)^2$ as a function of the dimensionless time $\chi t$ for various amount of phase fluctuations: no fluctuations(blue, solid); $D=10^{-5}\chi$~(green, dashed); $D=10^{-4}\chi$~(orange, dot-dashed); $D=10^{-3}\chi$~(red, dotted).}
\end{figure}

Again following Ref.~\cite{Zubairy1983}, we consider phase fluctuations $\delta \phi(t)$ about the relative phase $\phi$, approximated as a phase diffusion process characterized by the correlation functions
\begin{eqnarray}
\langle\delta\phi(t)\rangle &=&  \langle\delta\dot{\phi}(t)\rangle = 0, \nonumber \\
\langle\delta\phi(t)\delta\phi(t^{\prime})\rangle &=& D(t+t^{\prime}-|t-t^{\prime}|),\nonumber  \\
\langle\delta\dot{\phi}(t)\delta\dot{\phi}(t^{\prime})\rangle &=& 2D\delta(t-t^{\prime}),
\end{eqnarray}
where $D$ is the phase diffusion coefficient. From Ref.~\cite{Wodkiewicz1978}, (see details in Appendix A) we find
the variances of the quadratures
\begin{widetext}
\begin{eqnarray}
(\Delta\hat{X}_1(t))^{2} = \frac{1}{4}\left[\cosh{(2\chi t)}\exp(-Dt/2)-\sinh{(2\chi t)}\exp(-5Dt/3)\right], \\
(\Delta\hat{X}_2(t))^{2} = \frac{1}{4}\left[\cosh{(2\chi t)}\exp(-Dt/2)+\sinh{(2\chi t)}\exp(-5Dt/3)\right],
\end{eqnarray}
\end{widetext}
which reduce to
\begin{eqnarray}
&&(\Delta\hat{X}_1(t))^{2} = \frac{7}{48}Dt \exp(2\chi t) +\frac{1}{4}\exp(-2\chi t), \nonumber \\
&&(\Delta\hat{X}_2(t))^{2} = \frac{1}{4}\exp(2\chi t) + \frac{7}{48}Dt \exp(-2\chi t),
\end{eqnarray}
for $D t \ll 1$. Fig.~\ref{fig:phase} depicts the squared quadrature $(\Delta \hat{X}_1)^2$ as a function of the dimensionless time $\chi t$ for several values of the diffusion coefficient $D$, illustrating the disappearance of squeezing for long enough times, an effect familiar from quantum optics. 

\subsection{Clamping noise}
We now turn to a discussion of the effects of clamping noise on cantilever squeezing. We describe the thermal fluctuations coupled into the cantilever via a standard input-output formalism \cite{Milburn}, resulting in the Heisenberg-Langevin equations
\begin{equation}
\dot{\hat{b}} = -\chi \hat{b}^{\dagger} -\frac{\gamma}{2}\hat{b}+ \sqrt{\gamma}\hat{b}_{\rm in}.
\end{equation}
Here $\gamma$ is the damping rate of the cantilever,  and $\hat{b}_{\rm in}$ a noise operator that accounts for thermal fluctuations, with
\begin{eqnarray}
\langle\hat{b}_{\rm in}(t)\rangle&=&\langle\hat{b}^{\dagger}_{\rm in}(t)\rangle =0, \nonumber \\
\langle\hat{b}^{\dagger}_{\rm in}(t)\hat{b}_{\rm in}(t')\rangle &=& \bar{n}_{\rm th}\delta(t-t'), \nonumber  \\
\langle\hat{b}_{\rm in}(t)\hat{b}^{\dagger}_{\rm in}(t')\rangle &=& \left(\bar{n}_{\rm th}+1\right)\delta(t-t'), \nonumber \\
\langle\hat{b}^{\dagger}_{\rm in}(t)\hat{b}^{\dagger}_{\rm in}(t')\rangle &=&\langle\hat{b}_{\rm in}(t)\hat{b}_{\rm in}(t')\rangle =0,
\end{eqnarray}
and $\bar{n}_{\rm th}=k_BT/\hbar \omega_c'$.
This yields the quadrature evolution equations
\begin{eqnarray}
\frac{d}{dt}(\Delta\hat{X}_1)^{2} &=& -(2\chi+\gamma)(\Delta\hat{X}_1)^{2} +\frac{\gamma}{4}\left(2\bar{n}_{\rm th}+1\right), \\
\frac{d}{dt}(\Delta\hat{X}_2)^{2} &=& (2\chi-\gamma)(\Delta\hat{X}_2)^{2} +\frac{\gamma}{4}\left(2\bar{n}_{\rm th}+1\right),
\end{eqnarray}
and for a cantilever initially prepared in its ground state of center-of-mass motion, 
\begin{eqnarray}
(\Delta\hat{X}_1(t))^2 &=& \frac{(\chi/\gamma)-\bar{n}_{\rm th}}{2[1+2(\chi/\gamma)]}\exp[-(2\chi+\gamma)t] \nonumber \\
&+&\frac{1+2\bar{n}_{\rm th}}{4[1+2(\chi/\gamma)]}, \\
(\Delta\hat{X}_2(t))^2 &=& \frac{-(\chi/\gamma)-\bar{n}_{\rm th}}{2[1-2(\chi/\gamma)]}\exp[(2\chi-\gamma)t] \nonumber \\
&+&\frac{1+2\bar{n}_{\rm th}}{4[1-2(\chi/\gamma)]}.
\end{eqnarray}

\begin{figure}[t]
\includegraphics[width=0.48 \textwidth]{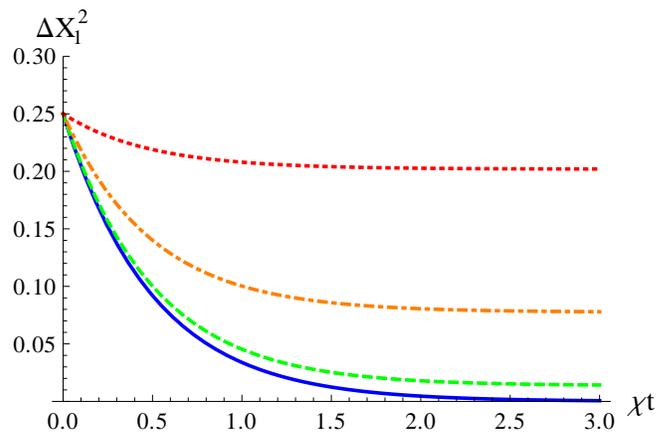}
\caption{\label{fig:quantum}(Color Online) $(\Delta X_1)^2$ as a function of scaled time with thermal fluctuations for different damping constants and occupation numbers: no fluctuations (blue, solid); $n_{\rm th} = 5, \gamma=10^{-2}\chi$ (green, dashed); $n_{\rm th} = 10, \gamma=3\times 10^{-2}\chi$ (orange, dot-dashed); and $n_{\rm th}=10, \gamma=8\times 10^{-2}\chi$ (red, dotted).}
\end{figure}

For $t \to \infty$, $(\Delta\hat{X}_1)^2$ approaches a steady state with reduced squeezing, as expected. Figure~\ref{fig:quantum} shows $(\Delta\hat{X}_1)^2$ for several thermal occupations of the clamp and two values of the mechanical damping rate $\gamma$. Consider for example a cantilever with a quality factor $Q_m= \omega_c'/2\gamma = 10^{4}$ , $\gamma \approx 2\pi\times 3.6$ $\mathrm{kHz}$ in a cryogenic environment at $T = 20$ mK. The phonon occupation number at $\omega_c'$ is  $\bar{n}_{\rm th} \simeq 5.7$. For a strong squeezing coupling constant of $\chi = 2\pi\times 2.8$ $\mathrm{MHz}$, this corresponds to a squeezing of $2 \times 10^{-3}$(21dB).

\section{Detection}

The detection of motional squeezing could be performed along the same lines as in the experiments of Ref.~\cite{rugar}, which generated classical squeezing in a parametrically driven mechanical cantilever and characterized it by measuring the two quadratures of oscillations using a fiber-optical interferometer. A more ambitious approach that offers many potential advantages \cite{pulsedQOM} involves a full determination of the cantilever state, rather than its covariances only. One promising way to achieve this goal involves first transferring that state to an optical field, where detection techniques developed in quantum optics can then be applied.

Quantum state transfer has already been the subject of a number of studies~\cite{Braunstein, TianWang, Khalili, squeezmech}. In particular, it is known that the two-mode beam-splitter interaction yields exact state transfer between two harmonic oscillators for appropriate interaction times and in the absence of dissipation, and that this type of interaction can be realized in principle in the optomechanical interaction between the harmonically bound end-mirror of a Fabry-P{\'e}rot and a near resonant intracavity light field.

To implement that detection scheme, we expand the scheme of Fig.~1 to couple the cantilever to the intracavity light field of a resonator whose moving end-mirror is attached to the cantilever. This could be achieved e.g. with a cantilever of the type used in nanoscale magnetic resonance imaging \cite{RugarPNAS}, see Fig.~5. Combined with a fixed large mirror, this can form a high finesse optomechanical resonator. Alternatively, one could also consider an arrangement where the cantilever forms a moving plate of a capacitor in a driven microelectronic LC-circuit~\cite{schwab,cool2}. In which case the coupling could actually be stronger, but one would be confronted with the lack of single-photon detectors in the microwave regime. 

In this section we consider two possible scenarios: In the first one the coupling to the optical field is present at all times, while in the second one a squeezed state of the cantilever is first prepared via magnetic dipole coupling to the classical tuning fork, and an optical field is subsequently turned on. We show that in both approaches the additional dissipation channel associated to the finite transmission of the Fabry-P\'erot leads to a significant reduction in squeezing and imperfect state transfer.

\begin{figure}[t]
\includegraphics[width=8cm]{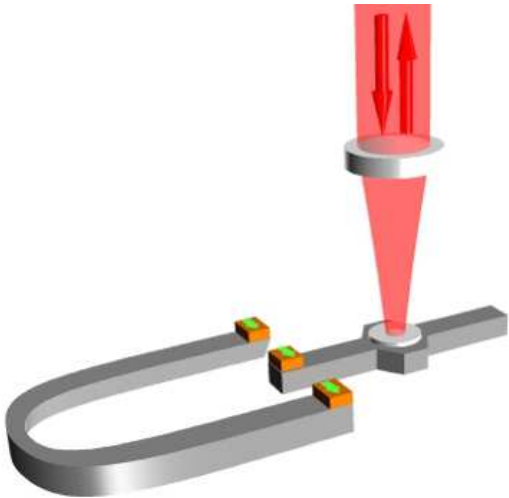}
\caption{\label{fig:system2} Schematics of the intracavity optical field optomechanically coupled to the magnetically driven cantilever. }
\end{figure}

\subsection{Continuous optical coupling}

With the additional optomechanical coupling of the cantilever to the optical field, the system Hamiltonian becomes
\begin{equation}
H_\mathrm{\rm sys} = H_{\rm c}+H_{\rm p}+H_{\rm om}+H_{\rm m}+H_{\gamma}+H_{\kappa}.
\label{Htotal}
\end{equation}
Here
\begin{equation}
H_{\rm c}=\hbar\omega_0\hat{a}^{\dagger}\hat{a}
\end{equation}
describes the optical cavity mode $\hat{a}$ of frequency $\omega_0$,
\begin{equation}
H_{\rm p}=i\hbar\eta(\hat{a}^{\dagger}e^{-i\omega_\mathrm{L}t}-\hat{a}e^{i\omega_\mathrm{L}t})
\end{equation}
accounts for the driving of the cavity by an external field of frequency $\omega_L$ at rate $|\eta|=\sqrt{2P\kappa/\hbar\omega_L}$, with $P$ the input power and $\kappa$ the cavity linewidth, and
\begin{equation}
H_\mathrm{om}=-\hbar g\hat{a}^{\dagger}\hat{a}(\hat{b}+\hat{b}^{\dagger})
\end{equation}
is the optomechanical coupling between the intracavity field and the cantilever mirror, with single-photon coupling frequency $g$. Finally,  $H_{\gamma}$ and $H_{\kappa}$ describe the interaction of the mirror and cavity field to thermal reservoirs and account for dissipation at rates $\gamma$ and $\kappa$, respectively, and $H_\mathrm{m}$ is the Hamiltonian of the magnetically driven cantilever of the previous sections.

Considering a coherent pump of constant amplitude red-detuned from the cavity resonance by $\omega_c'$, we expand the amplitudes of both the cavity field and cantilever oscillations as the sum of their expectation value and quantum fluctuations, 
\begin{eqnarray}
\hat{a} &=& \langle\hat{a}\rangle + \delta\hat{a} \equiv E_0 +  \delta\hat{a} , \nonumber \\
\hat{b} &=& \langle\hat{b}\rangle + \delta\hat{b}, 
\label{decomp}
\end{eqnarray}
where $\langle \delta \hat a\rangle = \langle \delta \hat b \rangle = 0$ and we neglect contributions from $\delta\hat{a}^\dag\delta\hat{a}$ compared to those from $\langle\hat{a}\rangle\delta\hat{a}^\dag+\langle\hat{a}\rangle^{*}\delta\hat{a}$ as usual, so that $|E_0|^2= \langle \hat a ^\dagger\rangle \langle \hat  a \rangle \approx \langle \hat a ^\dagger \hat a \rangle$. (Note that $E_0$ is dimensionless.) As is well known, this decomposition allows one to separate the optomechanical interaction Hamiltonian into a classical Kerr-type contribution proportional to $|E_0|^2$ and a beam-splitter interaction, so that the cantilever dynamics is approximately described by the Hamiltonian
\begin{eqnarray}
H &=& -\hbar g\left(E_0^{*}\delta\hat{a}\delta\hat{b}^{\dagger}+E_0\delta\hat{a}^{\dagger}\delta\hat{b}\right) +H_{\kappa} \nonumber\\	
&+&\frac{\hbar\chi}{2}\left(\delta\hat{b}^{\dagger 2}e^{-i\phi} + \delta\hat{b}^2e^{i\phi}\right)+H_{\gamma},
\end{eqnarray}
where $E_0$ accounts consistently for the Kerr nonlinearity. In steady state, the main consequence of the Kerr effect is a slight shift in the cavity resonance, an effect that can lead under appropriate conditions to optical bistability~\cite{Dorsel}. Away from this multistable regime, the intracavity amplitude $\langle \hat a \rangle \equiv E_0$ of the radiation field is given approximately by (see Appendix B) 
\begin{equation}
E_0 = \frac{\eta}{\kappa/2-i\left[\Delta+2g^2|E_0|^2/\omega_c^{\prime} \right]}.
\end{equation}
In the interaction picture, in the rotating wave approximation, and for  $\omega_c^{\prime} = -\Delta$, the equations of motion for $\delta\hat{a}$ and $\delta\hat{b}$ are, (see Eqs.~(\ref{adot}) and (\ref{bdot}))
\begin{eqnarray}
\dot{\delta\hat{a}} &=& -\frac{\kappa}{2}\delta\hat{a}+igE_0\delta\hat{b}+\sqrt{\kappa}\hat{a}_{\rm in}, \\
\dot{\delta\hat{b}} &=& -\frac{\gamma}{2}\delta\hat{b}+igE_0^{*}\delta\hat{a}
-\chi\delta\hat{b}^{\dagger} +\sqrt{\gamma}\hat{b}_{\rm in},
\end{eqnarray}
where we have assumed that
\begin{eqnarray}
\langle\hat{a}^{\dagger}_{\rm in}(t)\hat{a}_{\rm in}(t^{\prime})\rangle &=& 0, \nonumber \\
\langle\hat{a}_{\rm in}(t)\hat{a}^{\dagger}_{\rm in}(t^{\prime})\rangle &=& \delta(t-t^{\prime}),\nonumber \\
\langle\hat{a}_{\rm in}(t)\hat{a}_{\rm in}(t^{\prime})\rangle &=& 0.
\end{eqnarray}

We consider for concreteness the case $E_0=|E_0|e^{-i\pi/2}$, which holds when $\omega_c'\gg\kappa$. The equations of motion for the variances of the position quadrature of the optical and phonon modes are then given by
\begin{equation}
\dot{V} = A V+B,
\end{equation}
with
\begin{eqnarray}
A &=& \begin{bmatrix} -\kappa & 2g|E_0| & 0 \\ -g|E_0| & -\left(\kappa/2+\gamma/2+|\chi|\right) & g|E_0| \\ 0 & -2g|E_0| & -(2|\chi|+\gamma) \end{bmatrix}, \\
B &=& \left[ \begin{array}{c} \kappa/4 \\ 0 \\ (\gamma/8)(3n_{\rm th}+1) \end{array} \right], \\
V &=& \left[ \begin{array}{c} (\Delta\hat{X}_{\rm o,1})^{2} \\ (\Delta\hat{X}_{\rm om,1})^2 \\ (\Delta\hat{X}_{\rm m,1})^{2} \end{array} \right],
\end{eqnarray}
where $(\Delta\hat{X}_{i,1})^{2}$, $i \in \{\rm o, \rm om, \rm m \}$ are the variance of the position quadrature of the optical field, the covariance of position quadratures of the optical field and cantilever, and the variance of the position quadrature of the cantilever, respectively. 

\begin{figure}[t]
\includegraphics[width=0.48\textwidth]{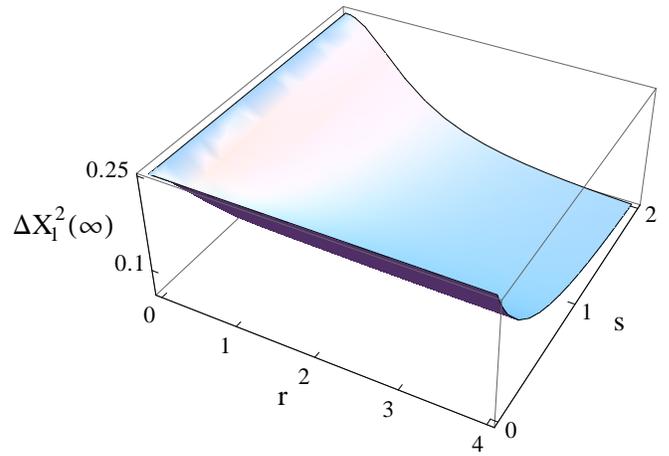}
\caption{Steady-state squeezing of the cavity field as a function of the dimensionless coupling parameters $r=g|E_0|/\kappa$ and $s=\chi/\kappa$.}
\label{fig:steadystatecav}
\end{figure}

\begin{figure}[t]
\includegraphics[width=0.48\textwidth]{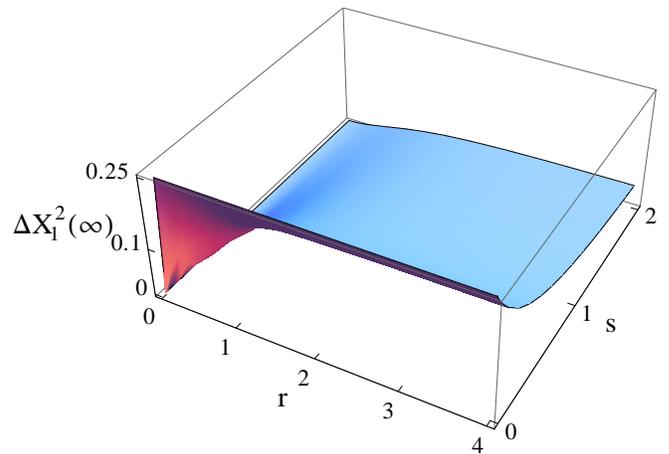}
\caption{Steady state squeezing of the cantilever as a function of the dimensionless coupling parameters $r=g|E_0|/\kappa$ and $s=\chi/\kappa$.}
\label{fig:steadystatecantilev}
\end{figure}

The steady-state squeezing of the intracavity field follows from some elementary algebra, but its general form is cumbersome and we omit it here. In the physically relevant regime $\kappa \gg \gamma$ it reduces to the simple form
\begin{equation}
(\Delta\hat{X}_{\rm o,1})^{2}(\infty)=\frac{2r^2+s+2s^2}{4(2r^2+s)(1+2s)},
\end{equation}
where $r = g|E_0|/\kappa$ is the classically amplified optomechanical coupling and $s=\chi/\kappa$. Figure~\ref{fig:steadystatecav} shows the steady-state squeezing of the optical field as a function of these parameters. It illustrates the monotonic increase in steady-state squeezing as $s$ and $r$ are increased, which is intuitively expected. However, this conclusion needs to be qualified by considering the steady-state limit of the cantilever squeezing. In the same limit it is likewise easily obtained as
\begin{equation}
(\Delta\hat{X}_{\rm m,1})^{2}(\infty)=\frac{r^2}{2(2r^2+s)(1+2s)},
\end{equation}
and is illustrated in Fig.~\ref{fig:steadystatecantilev} as a function of $r$ and $s.$ In contrast to the situation for the optical field, we observe now that while increasing  $s$ increases the degree of squeezing, as would be expected, increasing the optomechanical coupling results in a decrease in squeezing. Indeed, while in the absence of optomechanical coupling the state of the cantilever mode would be almost perfectly squeezed, this ceases to be the case once the optical coupling is present. However, realistic experimental parameters yield enough squeezing transfer to the cavity field to be successfully detected.

\begin{figure}[t]
\includegraphics[width=0.48\textwidth]{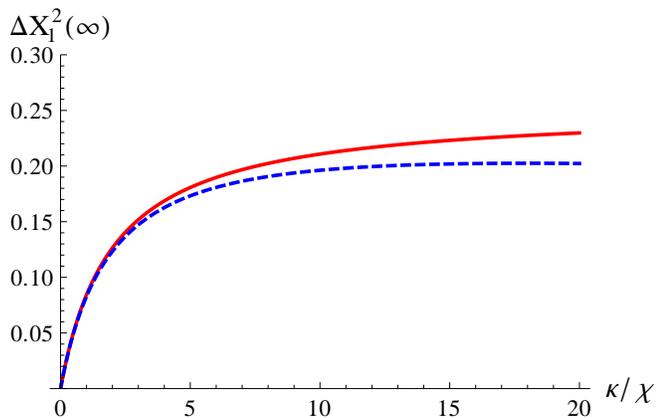}
\caption{(Color Online) Steady state squeezing of the cavity field (red, solid) and cantilever (blue, dashed) as a function of $\kappa/\chi$ in the resolved-side band regime for fixed $\chi$ and $g|E_0|/\chi = 9$.}
\label{fig:longtimekappa}
\end{figure}

The physical origin of this behavior is that in addition to the magnetic squeezing interaction the cantilever is now also subjected to the beam-splitter interaction. It results in a transfer of squeezing to the optical field, where it is now exposed to a dominating decoherence channel associated with the cavity loss rate $\kappa$, normally much faster than the mechanical decay rate $\gamma.$  This is more readily apparent in Fig.~\ref{fig:longtimekappa}, which clearly illustrates how cavity damping decreases the steady-state squeezing of both the cavity field and cantilever for fixed beam-splitter and squeezing coupling constants $g|E_0|$ and $\chi$.  

\subsection{Delayed detection}

The take-home message of the previous section is that while the coupling of the cantilever to the optical cavity allows detection of the squeezing, it does it at the cost of opening up a fast decoherence channel. For reasonable experimental parameters the resulting loss in squeezing is much larger than the limit imposed by thermal losses in the mechanics. This suggests that a better scenario might involve first preparing the cantilever in a strongly squeezed state, and only subsequently coupling it to the optical field. The issue with that approach is that it takes a time of the order of $\kappa^{-1}$ to switch on the intracavity optical field, a time during which the optical decoherence channel is already open. 

As before, we decompose the cantilever phonon field and intracavity optical field as the sum of their expectation value and quantum fluctuations, see Eqs.~(\ref{decomp}), except that $\langle a(t) \rangle \equiv E_0(t)$ is now an explicit function of time. The linearization process is questionable for very short times when the intracavity field is still extremely small. However, the optomechanical coupling is normally weak in that case, so that it should not qualitatively change the main features of the system dynamics.

As shown in Appendix B, Eqs.~(\ref{adot}) and (\ref{bdot}), the Heisenberg equations of motion for $\delta \hat a (t)$ and $\delta \hat b(t)$ are approximately given by
\begin{eqnarray}
\dot{\delta\hat{a}} &=& \left[i\Delta-\kappa/2\right]\delta\hat{a}+igE_0(t)(\delta\hat{b}+\delta\hat{b}^{\dagger})+\sqrt{\kappa}\hat{a}_{\rm in}, \\
\dot{\delta\hat{b}} &=& \left[-i\omega_c^{\prime}-\gamma/2\right]\delta\hat{b}+ig\left [E_0^*(t)\delta\hat{a}+E_0(t)\delta\hat{a}^{\dagger}\right ] \nonumber\\ 
&-& 4i\chi\cos(\omega_f t+ \phi)(\delta\hat{b}+\delta\hat{b}^{\dagger})+\sqrt{\gamma}\hat{b}_{\rm in},
\end{eqnarray}
and the evolution of $E_0(t)$ is determined by Eqs.~(\ref{at1})-(\ref{at2}). From these equations, it is possible to derive a closed set of equations for the first and second moments of the operators $\delta \hat a$ and $\delta \hat b$, see Appendix B. These equations could not be solved analytically, so this subsection presents selected numerical results that illustrate the main features of the system dynamics. 

\begin{figure}
\includegraphics[width=0.48 \textwidth]{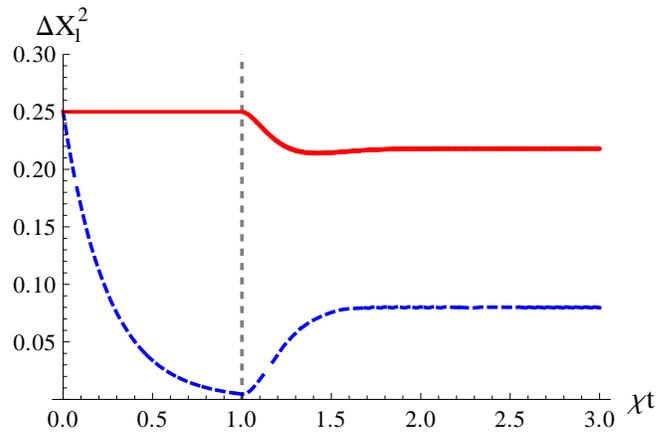}
\caption{(Color Online) Squared quadratures of position for the cavity field (red, solid) and the mechanical oscillator (blue, dashed) as a function of scaled time. The optomechanical coupling is turned on at the dimensionless time $\chi t=1$ and coherently builds up towards $g|E_0|/\chi=9, \kappa/\chi = 10$.}
\label{fig:transfersaturated}
\end{figure}

\begin{figure}
\includegraphics[width=0.48 \textwidth]{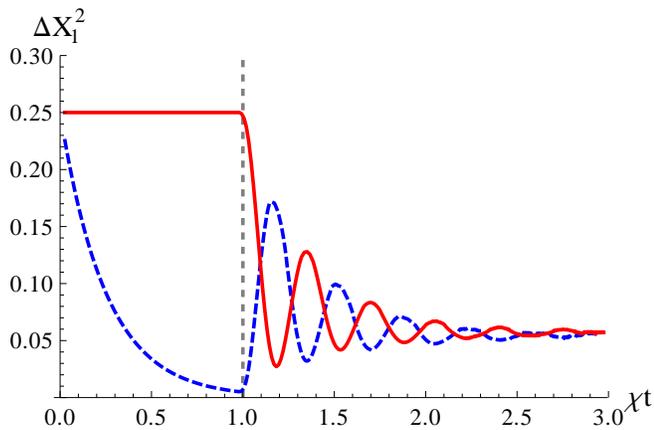}
\caption{(Color Online) Squared quadratures of position in the cavity field (red, solid) and the mechanical oscillator (blue dashed) with strong coherent optomechanical coupling. Here $g|E_0|/\chi=9, \kappa/\chi = 1$.}
\label{fig:transferstrongcoup}
\end{figure}

Fig.~\ref{fig:transfersaturated}, which is for a relatively high-loss optical cavity that allows for a fast switching of the optical field, shows the coupled dynamics of the cantilever and optical fields in a situation where the cantilever was first prepared in a squeezed state, before the optical field is switched on at $t_0=\chi^{-1}$. It illustrates a situation where squeezing transfer suffers from the broad decoherence channel of the optical cavity. Thus the squeezing is not efficiently detectable in the cavity field. Note also that since the beam-splitter interaction frequency $g|E_0| \simeq \kappa$ in that example, the oscillatory coherent state transfer between the cantilever and the optical field is strongly suppressed, with the energy of the cantilever-field system being rapidly lost through the optical decay channel. A much more significant coherent exchange between the two subsystems requires either a stronger field amplitude $|E_0|$, or a slower decay of the light field, so that $g|E_0| \gg \kappa$. Such an example is illustrated in Fig. \ref{fig:transferstrongcoup}, which shows the characteristic coherent state transfer between the phonon and photon field, as expected. One problem here is of course that by decreasing the cavity damping rate, one requires a longer time to turn on the light field to its final value $|E_0|$, thereby increasing the role of dissipation. Still, in this situation it is possible to achieve a reasonably good transfer of squeezing from the cantilever to the optical field.

\begin{figure}
\includegraphics[width=0.48 \textwidth]{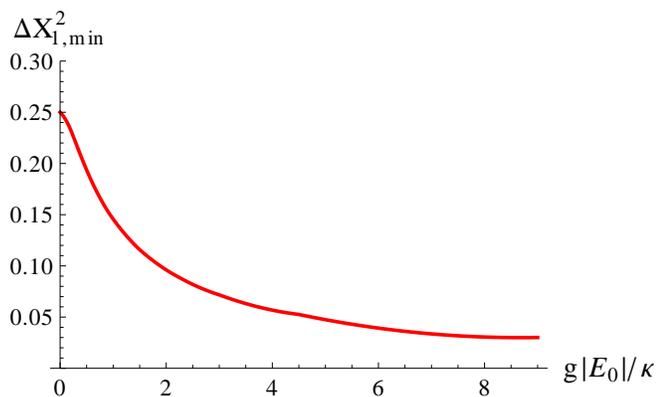}
\caption{Minimum values of squared quadrature variance in the cavity field plotted as a function of $g|E_0|/\kappa$.}
\label{fig:minimum}
\end{figure}

During the coherent state transfer, the maximum squeezing in the intracavity field occurs after half an exchange period. In Fig.~\ref{fig:minimum}, the minimum values of the quadrature variance in the cavity field are plotted as a function of $g|E_0|/\kappa$. As is expected, smaller cavity damping rate or lager coupling strength gives rise to stronger maximum squeezing in the cavity field. 

\section{Conclusion}
In summary, we have presented a theoretical analysis of the motional squeezing of a cantilever magnetically coupled to a classical tuning fork via microscopic magnetic dipoles. We showed that this coupling can result in significant squeezing of a quadrature of motion of the cantilever if appropriately driven by a classical force, and found that the system is robust against various sources of noise, with phase noise in the driving of the classical driving tuning fork the dominant source of decoherence. We proposed a scheme for the detection of the effect based on state transfer to the intracavity field of an optical resonator with one end-mirror formed by the oscillating cantilever. Challenges to the measurement process associated with the additional decoherence channel opened by the coupling to the optical resonator were discussed. 

It has recently been proposed that pulsed optomechanical configurations permit mapping the quantum state of optomechanical oscillators by using a sequence of appropriately shaped optical pulses separated in time by half a vibration period of the mechanical system \cite{Vanner}. This approach presents several advantages, the first one being that it allows for the use of low finesse optical resonators that permit fast switching and the second being that the oscillator is coupled to the optical dissipation channel for very short times only. Unfortunately, this scheme relies on the mechanical oscillator being subject only to free evolution between the light pulses, which is not the case here since the squeezing interaction is acting at all times. It is not clear how it could be rapidly be switched off to make the pulsed detection scheme applicable. Future studies will consider whether adapting pulsed detection scenarios to the present situation may be possible. We will also consider the quantum dynamics of the tuning fork, the interaction between these two subsystems on a quantum level as well as further possibilities of optomechanical coupling to control and probe the system, including the use of multimode light fields.

\acknowledgments
This work is supported by the US National Science Foundation, the DARPA ORCHID and QuASAR programs,  and the US Army Research Office.

\appendix
\section{Effect of phase fluctuations}
This appendix presents details of the evaluation of the effect of phase fluctuations for the case of an asymmetric setup. The symmetric case is analogous. In the presence of phase fluctuations, the Heisenberg evolution of the phonon annihilation operator $\hat b(t)$ becomes
\begin{equation}
\frac{d}{dt} {\hat b}(t) = -\chi e^{-i \delta \phi(t)} {\hat b}^\dagger(t).
\end{equation}
and the full dynamics of the system can be expressed in the general form
\begin{equation}
\frac{dY}{dt} = \left(A + i\delta\dot{\phi}(t)B\right)Y,
\end{equation}
where $Y$ is the vector of bilinear operators $\hat{b}^{2}$, $\hat{b}^{\dagger 2}$, and $\hat{b}^{\dagger}\hat{b}+\hat{b}\hat{b}^{\dagger}$. Due to the form of the Heisenberg equations of motion for the problem at hand, the phase fluctuations $\delta\phi(t)$ can only be eliminated in one bilinear entry in $Y$, and it is therefore impossible to readily perform the statistical average over phase noise. As shown by Wodkievicz \cite{Wodkiewicz1978} this difficulty can be circumvented by solving two systems of matrix equations separately for two particular choices of $Y$, $\hat{b}^{2}$ and $\hat{b}^{\dagger}\hat{b}+\hat{b}\hat{b}^{\dagger}$.
For $\hat{b}^{2}$ we have
\begin{equation}
Y = \left[ \begin{array}{c} \hat{b}^{2} \\ e^{-i\delta\phi(t)}(\hat{b}\hat{b}^{\dagger}+\hat{b}^{\dagger}\hat{b}) \\ e^{-2i\delta\phi(t)}\hat{b}^{\dagger2} \end{array} \right],
\end{equation}
\begin{equation}
A = \begin{bmatrix} 0 & -\chi & 0 \\ -2\chi & 0 & -2\chi \\ 0 & -\chi & 0 \end{bmatrix},
\end{equation}
\begin{equation}
B = \begin{bmatrix} 0 & 0 & 0 \\ 0 & -1 & 0 \\ 0 & 0 & -2 \end{bmatrix},
\end{equation}
while for  $\hat{b}^{\dagger}\hat{b}+\hat{b}\hat{b}^{\dagger}$
\label{bdagb+hc}
\begin{equation}
Y = \left[ \begin{array}{c} \hat{b}\hat{b}^{\dagger}+\hat{b}^{\dagger}\hat{b} \\ e^{i\delta\phi(t)}\hat{b}^{2} \\ e^{-i\delta\phi(t)}\hat{b}^{\dagger2} \end{array} \right],
\end{equation}
\begin{equation}
A = \begin{bmatrix} 0 & -2\chi & -2\chi \\ -\chi & 0 & 0 \\ -\chi & 0 & 0 \end{bmatrix},
\end{equation}
\begin{equation}
B = \begin{bmatrix} 0 & 0 & 0 \\ 0 & 1 & 0 \\ 0 & 0 & -1 \end{bmatrix}.
\end{equation}
If the initial state is uncorrelated with the fluctuations, as is physically the case, this form of equations can be solved exactly to give
\begin{equation}
\langle Y(t)\rangle = e^{(A-DB^{2})t}\langle Y(0)\rangle,
\label{E1}
\end{equation}
so that
\begin{eqnarray}
&&\langle\hat{b}^{2}\rangle \approx -\frac{1}{2}\sinh{(2\chi t)}\exp[-5Dt/3], \\
&&\langle\hat{b}^{\dagger}\hat{b}+\hat{b}\hat{b}^{\dagger}\rangle \approx \cosh{(2\chi t)}\exp[-Dt/2],
\end{eqnarray}
where we have assumed that $D \ll \chi$ for simplicity.

\section{Equations of motion for optically coupled system}

In the rotating frame at frequency $\omega_L$ the system Hamiltonian~(\ref{Htotal}) is 
\begin{eqnarray}
H &=& -\hbar\Delta \hat{a}^{\dagger}\hat{a}-\hbar g\hat{a}^{\dagger}\hat{a}(\hat{b}+\hat{b}^{\dagger})+i\hbar\eta(\hat{a}^{\dagger}-\hat{a})+\hbar\omega_c^{\prime}\hat{b}^{\dagger}\hat{b} \nonumber\\ 
&+&\hbar\chi\cos(\omega_f t+\phi)(\hat{b}+\hat{b}^{\dagger})^2+H_{\gamma}+H_{\kappa},
\end{eqnarray}
where 
\begin{equation}
\Delta = \omega_L-\omega_0,
\end{equation}
and the  equations of motion for $\hat{a}$ and $\hat{b}$ are
\begin{eqnarray}
\dot{\hat{a}} &=& i\Delta\hat{a}+ig(\hat{b}+\hat{b}^{\dagger})\hat{a}-\frac{\kappa}{2}\hat{a}+\eta+\sqrt{\kappa
}\hat{a}_{\rm in}, \\
\dot{\hat{b}} &=& -i\omega_c^{\prime}\hat{b}+ig\hat{a}^{\dagger}\hat{a} \nonumber\\ 
&-& 2i\chi\cos(\omega_f t+\phi)(\hat{b}+\hat{b}^{\dagger}) -\frac{\gamma}{2}\hat{b}+ \sqrt{\gamma}\hat{b}_{\rm in},
\end{eqnarray}
where $\phi$ is the phase difference between the tuning fork and cantilever. 
Introducing $\delta\hat{a} = \hat{a}-\langle\hat{a}\rangle$ and $\delta\hat{b} = \hat{b}-\langle\hat{b}\rangle$ yields then for $\delta\hat{a}$ and $\delta\hat{b}$ the linearized equations of motion
\begin{eqnarray}
\label{adot}
\dot{\delta\hat{a}} &\approx & \left[i\Delta -\kappa/2\right]\delta\hat{a}+ig \langle \hat b + \hat b^\dagger \rangle \delta \hat a  \nonumber \\
&+&ig\langle\hat{a}\rangle(\delta\hat{b}+\delta\hat{b}^{\dagger})+ \sqrt{\kappa}\hat{a}_{\rm in} \\
&\approx& \left[i\Delta -\kappa/2\right]\delta\hat{a} + ig\langle\hat{a}\rangle(\delta\hat{b}+\delta\hat{b}^{\dagger})+ \sqrt{\kappa}\hat{a}_{\rm in}, \nonumber
\end{eqnarray}
\begin{eqnarray}
\label{bdot}
\dot{\delta\hat{b}} &\approx & \left[-i\omega_c^{\prime}-\gamma/2\right]\delta\hat{b}+ig(\langle\hat{a}\rangle^{*}\delta\hat{a}+\langle\hat{a}\rangle\delta\hat{a}^{\dagger}) \nonumber\\ 
&-& 2i\chi\cos(\omega_f t+ \phi)(\delta\hat{b}+\delta\hat{b}^{\dagger})+\sqrt{\gamma}\hat{b}_{\rm in}.
\end{eqnarray} 
The second, approximate form of Eq.~(\ref{adot}) results from the fact that the mean phonon number is of order unity, which is much smaller than the mean number of intractivity photons ($10^4 - 10^8$). Under these conditions one can neglect the term $ig \langle \hat b + \hat b^\dagger \rangle$ in that equation.

From these approximate linearized equations we easily obtain the equations of motion for expectation values of the quadrature operators of $\hat{a}$ and $\hat{b}$, 
\begin{eqnarray}
\langle\dot{\hat{X_a}}\rangle &=& -\Delta \langle\hat{Y_a}\rangle-\frac{\kappa}{2}\langle\hat{X_a}\rangle-2g\langle\hat{X_b} \hat{Y_a}\rangle + \eta, \\
\langle\dot{\hat{Y_a}}\rangle &=& \Delta \langle\hat{X_a}\rangle-\frac{\kappa}{2}\langle\hat{Y_a}\rangle+2g\langle\hat{X_b} \hat{X_a}\rangle, \\
\langle\dot{\hat{X_b}}\rangle &=& \omega_c^{\prime}\langle\hat{Y_b}\rangle -\frac{\gamma}{2}\langle\hat{X_b}\rangle, \\
\langle\dot{\hat{Y_b}}\rangle &=& -\left [\omega_c^{\prime} + 4\chi\cos(\omega_f t +\phi) \right ] \langle\hat{X_b}\rangle \nonumber \\
&+&g\langle\hat{N}_a\rangle -\frac{\gamma}{2}\langle\hat{Y_b}\rangle, \\
\langle\dot{\hat{N_a}}\rangle &=& 2\eta\langle\hat{X}_a\rangle -\kappa\langle\hat{N}_a\rangle,
\end{eqnarray}
where $\hat{X}_{a}=(\hat{a}+\hat{a}^{\dagger})/2$, $\hat{Y}_{a}=(\hat{a}-\hat{a}^{\dagger})/2i$, and  $\hat{N_a}$ is the intracavity photon number operator. In the regime where $\omega_c^{\prime} \gg \chi \gg \gamma$  and $\langle\hat{N_a}\rangle \equiv |E_0|^2 \gg 1$, the change in the classical component of the intracavity field due the cantilever oscillations remains very small. It can be ignored in determining the dynamics of the beam-splitter coupling constant  $g|E_0(t)|$, which is then governed by the approximate equations of motion
 \begin{eqnarray}
\langle\dot{\hat{X_a}}\rangle &=& -\Delta \langle\hat{Y_a}\rangle-\frac{\kappa}{2}\langle\hat{X_a}\rangle-2g\langle\hat{X_b} \hat{Y_a}\rangle + \eta,\label{at1} \\
\langle\dot{\hat{Y_a}}\rangle &=& \Delta \langle\hat{X_a}\rangle-\frac{\kappa}{2}\langle\hat{Y_a}\rangle+2g\langle\hat{X_b} \hat{X_a}\rangle, \\
\langle\dot{\hat{X_b}}\rangle &=& \omega_c^{\prime}\langle\hat{Y_b}\rangle, \\
\langle\dot{\hat{Y_b}}\rangle &=& -\omega_c^{\prime}\langle\hat{X_b}\rangle +g|E_0|^2, \\
\dot{ |E_0|^2} &=& 2\eta\langle\hat{X}_a\rangle -\kappa |E_0|^2.\label{at2}
\end{eqnarray}
These equations yield the time evolution of $\langle\hat{a}(t)\rangle$, as well as its classical steady-state value. We find, upon factorizing $\langle\hat{X_b} \hat{Y_a}\rangle$,
\begin{equation}
|E_0|_{\rm ss} = \frac{\eta}{\kappa/2 -i\left[\Delta_{\rm eff} + 2g^2\langle|E_0|_{\rm ss}^2 \rangle/\omega_c^{\prime}\right]}.
\end{equation}

In the rotating frame at frequency $\omega_L$, the position quadratures of both cantilever and cavity field are defined as
\begin{eqnarray}
\hat{X}_{1, \rm o} = \frac{1}{2}(\delta\hat{a}e^{-i\Delta t}+\delta\hat{a}^{\dagger}e^{i\Delta t}), \\
\hat{X}_{1, \rm m} = \frac{1}{2}(\delta\hat{b}e^{i\omega_c^{\prime} t}+\delta\hat{b}^{\dagger}e^{-i\omega_c^{\prime} t}),
\end{eqnarray}
In order to calculate their variances we need to have the expectation value of second moments of the fluctuations. Taking quantum averages of the equations of motion for these quantities results in the closed set of equations
\begin{widetext}
\begin{eqnarray}
\frac{d}{dt}\langle\delta\hat{a}^2\rangle &=& 2i\Delta\langle\delta\hat{a}^2\rangle-\kappa\langle\delta\hat{a}^2\rangle
+2ig\langle\hat{a}\rangle\langle\delta\hat{a}\delta\hat{b}+\delta\hat{a}\delta\hat{b}^{\dagger}\rangle, \\
\frac{d}{dt}\langle\delta\hat{a}^{\dagger}\delta\hat{a}\rangle &=& -\kappa\langle\delta\hat{a}^{\dagger}\delta\hat{a}\rangle-ig\left[\langle\hat{a}\rangle^{*}\langle\delta\hat{a}\delta\hat{b}+\delta\hat{a}\delta\hat{b}^{\dagger}\rangle -\langle\hat{a}\rangle\langle\delta\hat{a}^{\dagger}\delta\hat{b} +\delta\hat{a}^{\dagger}\delta\hat{b}^{\dagger}\rangle\right], \\
\frac{d}{dt}\langle\delta\hat{b}^2\rangle  &=& -2i\omega_c^{\prime}\langle\delta\hat{b}^2\rangle -\gamma\langle\delta\hat{b}^2\rangle +2ig\left[\langle\hat{a}\rangle^{*}\langle\delta\hat{a}\delta\hat{b}\rangle +\langle\hat{a}\rangle\langle\delta\hat{a}^{\dagger}\delta\hat{b}\rangle\right] \nonumber \\
&-&2i\chi\cos(\omega_f t+ \phi)\langle2\delta\hat{b}^2+\delta\hat{b}^{\dagger}\delta\hat{b}+\delta\hat{b}\delta\hat{b}^{\dagger}\rangle, \\
\frac{d}{dt}\langle\delta\hat{b}^{\dagger}\delta\hat{b}\rangle &=& -\gamma\langle\delta\hat{b}^{\dagger}\delta\hat{b}\rangle-ig\left[\langle\hat{a}\rangle^{*}\langle\delta\hat{a}\delta\hat{b}-\delta\hat{a}\delta\hat{b}^{\dagger}\rangle +\langle\hat{a}\rangle\langle\delta\hat{a}^{\dagger}\delta\hat{b}-\delta\hat{a}^{\dagger}\delta\hat{b}^{\dagger}\rangle \right] \nonumber \\
&+&2i\chi\cos(\omega_f t+ \phi)\langle\delta\hat{b}^2-\delta\hat{b}^{\dagger 2}\rangle +\gamma n_{\rm th},\\
\frac{d}{dt}\langle\delta\hat{a}\delta\hat{b}\rangle &=& i(\Delta-\omega_c^{\prime})\langle\delta\hat{a}\delta\hat{b}\rangle-\left[(\kappa + \gamma)/2\right]\langle\delta\hat{a}\delta\hat{b}\rangle +ig\left[\langle\hat{a}\rangle\langle\delta\hat{b}^2+\delta\hat{b}^{\dagger}\delta\hat{b}\rangle +\langle\hat{a}\rangle^{*}\langle\delta\hat{a}^2\rangle +\langle\hat{a}\rangle\langle\delta\hat{a}\delta\hat{a}^{\dagger}\rangle\right] \nonumber \\
&&-2i\chi\cos(\omega_f t+ \phi)\langle\delta\hat{a}\delta\hat{b}+\delta\hat{a}\delta\hat{b}^{\dagger}\rangle, \\
\frac{d}{dt}\langle\delta\hat{a}\delta\hat{b}^{\dagger}\rangle &=&
i(\Delta + \omega_c^{\prime})\langle\delta\hat{a}\delta\hat{b}^{\dagger}\rangle -\left[(\kappa + \gamma)/2 \right ] \langle\delta\hat{a}\delta\hat{b}^{\dagger}\rangle +ig\left[\langle\hat{a}\rangle\langle\delta\hat{b}^{\dagger2} + \delta\hat{b}\delta\hat{b}^{\dagger}\rangle -\langle\hat{a}\rangle^{*}\langle\delta\hat{a}^{2}\rangle - \langle\hat{a}\rangle\langle\delta\hat{a}\delta\hat{a}^{\dagger}\rangle\right] \nonumber \\
&+&2i\chi\cos(\omega_f t+ \phi)\langle\delta\hat{a}\delta\hat{b}+\delta\hat{a}\delta\hat{b}^{\dagger}\rangle, \\
\end{eqnarray}
\end{widetext}
These are the equations that are solved numerically to obtain the figures of section IV.B.

\end{document}